\documentclass[sigconf,nonacm]{acmart}
\AtBeginDocument{%
  }

\setcopyright{none}
\usepackage{balance}
\usepackage{float}
\usepackage{pifont}
\newcommand{\cmark}{\ding{51}} %√
\usepackage{fontawesome5}
\usepackage{xcolor}
\newcommand{\first}{\textcolor{yellow!70!orange}{\faMedal}}
\newcommand{\second}{\textcolor{gray}{\faMedal}}
\newcommand{\third}{\textcolor{brown}{\faMedal}}
\usepackage{booktabs}
\usepackage{caption}
\usepackage[table]{xcolor}
\definecolor{lightblue}{RGB}{232,242,255}

\begin{document}

\title{Semantic-Enhanced Sparse Temporal Modeling for Dynamic Engagement Estimation in Multimodal Interactions}


% \author{Zhenzhen Hu}
% \affiliation{%
%   \institution{School of Computer Science and Information Engineering, Hefei University of Technology}
%   % \streetaddress{1 Th{\o}rv{\"a}ld Circle}
%   \city{Hefei}
% \email{zzhu@hfut.edu.cn}

\author{Yangchen Yu}
\affiliation{%
  \institution{Hefei University of Technology}
  % \streetaddress{1 Th{\o}rv{\"a}ld Circle}
  \city{Hefei}
  \country{China}}
\email{shijie@mail.hfut.edu.cn}

\author{Taoliang Wang}
\affiliation{%
  \institution{Hefei University of Technology}
  % \streetaddress{1 Th{\o}rv{\"a}ld Circle}
  \city{Hefei}
  \country{China}}
\email{hhhwtlww@gmail.com}

\author{Jia Li}
\authornote{Corresponding author.}
\affiliation{%
  \institution{Hefei University of Technology}
  % \streetaddress{1 Th{\o}rv{\"a}ld Circle}
  \city{Hefei}
  \country{China}}
\email{jiali@hfut.edu.cn}

\author{Nor Mazlina Ghazali}
\affiliation{%
  \institution{Universiti Malaysia Sarawak}
  \city{Kota Samarahan}
  \state{Sarawak}
  \country{Malaysia}}
\email{gnmazlina@unimas.my}

\author{Mohd Nazim Mohtar}
\affiliation{%
  \institution{Universiti Putra Malaysia}
  \city{Serdang}
  \state{Selangor}
  \country{Malaysia}}
\email{nazim@upm.edu.my}

\author{Bingkun Bao}
\affiliation{%
  \institution{Hefei University of Technology}
  % \streetaddress{1 Th{\o}rv{\"a}ld Circle}
  \city{Hefei}
  \country{China}}
\email{bingkunbao@hfut.edu.cn}

\author{Richang Hong}
\affiliation{%
  \institution{Hefei University of Technology}
  % \streetaddress{1 Th{\o}rv{\"a}ld Circle}
  \city{Hefei}
  \country{China}}
\email{hongrc.hfut@gmail.com}

\renewcommand{\shortauthors}{Yu et al.}

\begin{abstract}
Dynamic engagement estimation aims to infer participants’ evolving engagement states from continuous multimodal interaction streams. Existing approaches typically combine acoustic, pose, and visual representations with temporal models to capture observable interaction patterns. However, conversational semantics are difficult to align with frame-level engagement annotations in the same manner as continuous audio-visual signals, leaving dialogue context less effectively exploited in existing temporal modeling pipelines. Moreover, engagement annotations often evolve smoothly over time, making dense frame-level modeling computationally redundant and potentially sensitive to local annotation noise. To address these issues, we propose \textbf{SEST}, a \textbf{S}emantically-\textbf{E}nhanced \textbf{S}parse \textbf{T}emporal framework that models engagement-related information from both semantic and temporal perspectives. SEST leverages Multimodal Large Language Models (MLLMs) to extract multimodal semantic embeddings, providing dialogue-aware contextual information beyond conventional frame-level features. In parallel, SEST replaces dense-to-dense temporal modeling with a sparse-to-dense pathway, where continuous frame sequences are first encoded into compact temporal dynamics to reduce redundant computation and broaden temporal coverage, then fused with semantic context and recovered to the original label resolution for dense engagement prediction. Experiments on the  MultiMediate'26 Multi-domain Engagement Estimation (MEE2026) benchmark demonstrate the effectiveness of SEST, which achieved first place in the challenge\footnote{Code is available at \url{https://github.com/MSA-LMC/SEST}.}.

\end{abstract}

\begin{CCSXML}
<ccs2012>
   <concept>
       <concept_id>10010147.10010257</concept_id>
       <concept_desc>Computing methodologies~Machine learning</concept_desc>
       <concept_significance>500</concept_significance>
       </concept>
   <concept>
       <concept_id>10003120.10003121</concept_id>
       <concept_desc>Human-centered computing~Human computer interaction (HCI)</concept_desc>
       <concept_significance>300</concept_significance>
       </concept>
 </ccs2012>
\end{CCSXML}

\ccsdesc[500]{Computing methodologies~Machine learning}
\ccsdesc[300]{Human-centered computing~Human computer interaction (HCI)}

\keywords{Dynamic Engagement Estimation, Sparse Temporal Modeling, Multimodal Large Language Models, Multimodal Interaction}


\maketitle

\section{Introduction}

As a key indicator of interaction quality, engagement reflects participants' involvement and responsiveness during communication \cite{10.3389,sidner2002human}. Reliable estimation of such states is essential for socially intelligent systems in complex real-world interactions. Dynamic engagement estimation further requires tracking continuously evolving states from multimodal streams \cite{muller2022multimediate,reece2023candor,muller2023multimediate,withanage25_mm}, where observable behaviors and contextual factors are intertwined over time.

Current engagement estimation pipelines are largely built around frame-aligned multimodal signals \cite{yu2025generalizable,joshi2023multimodal,suzuki2025data}. Audio-visual features are naturally compatible with frame-level annotations and therefore serve as convenient inputs for temporal models. This design, however, also biases the modeling process toward locally observable behaviors \cite{inoue2018engagement}. In real interactions, conversational semantics and interaction context also play an important role in shaping engagement \cite{10.3389/fcomp.2023.1062342}, but they are difficult to align with frame-level audio-visual features and annotations. As a result, such high-level semantic information remains insufficiently explored in existing engagement estimation pipelines.

Another key challenge lies in the tension between long-range temporal perception and dense frame-level modeling. Reliable engagement estimation benefits from broader temporal context \cite{kumar2024towards, li2024dat, yu2023sliding,foster2017automatically}, as engagement often evolves gradually through behavioral trends and interaction changes. However, densely enlarging the temporal window introduces many repeated observations from adjacent frames, leading to higher computational cost and greater sensitivity to local annotation noise, especially under limited training data \cite{yu2023sliding,karimah2022automatic,alharbi2026hierarchical}. This may hinder the learning of robust and generalizable temporal dynamics, calling for a more compact temporal modeling strategy for continuous engagement prediction \cite{stappen2022estimation,tang2025asyrec}.

% To address these challenges, we propose \textbf{SEST}, a \textbf{S}emantically-\textbf{E}nhanced \textbf{S}parse \textbf{T}emporal framework for multimodal dynamic engagement estimation. To incorporate high-level conversational cues, SEST introduces \textbf{MLLM-based Semantic Enhancement Modeling}, extracting semantic embeddings from interaction segments to enrich frame-level behavioral representations with dialogue-aware contextual information. For temporal prediction, SEST further introduces \textbf{Sparse-to-Dense Temporal Module}, which reorganizes dense frame-level modeling into a compact temporal pathway for broader context perception and efficient continuous prediction. Combined with semantic enhancement, this design enables SEST to capture contextual semantics and temporal dynamics without excessive frame-level redundancy.

To address these challenges, we propose \textbf{SEST}, a \textbf{S}emantically-\textbf{E}nhanced \textbf{S}parse \textbf{T}emporal framework for multimodal dynamic engagement estimation. SEST introduces \textbf{MLLM-based Semantic Enhancement Modeling} to derive dialogue-aware semantic representations from interaction segments, and a \textbf{Sparse-to-Dense Temporal Module} that models compact temporal dynamics over broader contexts while recovering dense frame-level predictions. Together, the two components integrate high-level interaction semantics with compact temporal modeling.

The effectiveness of SEST is validated on the MEE2026 dynamic engagement estimation benchmark \cite{funk2026multimediate}, where our method achieved first place in the challenge. Our main contributions are summarized as follows:

\begin{enumerate}
\item \textbf{MLLM-based Semantic Enhancement Modeling} that leverages the semantic knowledge of Multimodal Large Language Models to derive high-level semantic embeddings, enabling dialogue-aware context to support dynamic engagement estimation.

% \item \textbf{Sparse-to-dense temporal Module} that reduces redundant frame-level computation while expanding the effective temporal perception range for continuous engagement prediction.
\item \textbf{Sparse-to-Dense Temporal Module} that reduces temporal sequence redundancy while expanding the effective temporal perception range for continuous engagement prediction.

\item \textbf{Best cross-dataset overall performance} across the MEE2026 benchmark, demonstrating the effectiveness and robustness of SEST under different evaluation settings.

\end{enumerate}

\section{Related Works}
{\bfseries Dynamic Multimodal Engagement Estimation.} Dynamic multimodal engagement estimation aims to track a participant's level of involvement through continuous behavioral signals, such as facial expressions, body gestures, and acoustic patterns \cite{chen2019faceengage,monkaresi2016automated,karimah2022automatic,yu2004detecting}. To capture these temporal dynamics, modern frameworks widely employ recurrent networks, sequence-to-sequence models \cite{yu2023sliding,huang2019fine}, or parallel attention architectures \cite{yu2025generalizable}. Since human engagement transitions are temporally smooth, uniform dense modeling of multimodal features inevitably introduces heavy information redundancy and heightens sensitivity to local tracking noise \cite{wang2016temporal,feichtenhofer2019slowfast,dutson2023eventful}. This motivates temporal modeling that reduces redundancy while preserving dense prediction \cite{wang2016temporal,tong2022videomae}. By sparsely sampling multimodal inputs and recovering dense representations prior to prediction, our approach reduces redundant temporal computations while preserving the continuous prediction resolution.  

% {\bfseries Semantic Representation via Multimodal Large Language Models.}
% Extracting high-level semantic representations is crucial for understanding complex human interactions, yet conventional engagement estimation methods heavily rely on low-level sensory features that lack contextual abstraction \cite{li2024dat,yu2023sliding}. To bridge this semantic gap, Multimodal Large Language Models (MLLMs) have emerged as powerful tools capable of generating rich semantic embeddings that capture deep cognitive and situational states \cite{zhang2026affective,li2025emoverse,lin-etal-2024-video,zhang-etal-2023-video}.Specifically, recent advanced studies have demonstrated that large models possess a strong capacity to analyze subtle human emotion preferences and parse high-level psychological nuances \cite{lian2025emoprefer,sabour-etal-2024-emobench}. Inspired by these advancements, we leverage window-level semantic representations from MLLMs to capture the global interactive context\cite{maaz-etal-2024-video}. By integrating these high-level semantic cues with local temporal behaviors, our approach achieves a more comprehensive and context-aware engagement prediction.

{\bfseries Semantic Representation via Multimodal Large Language Models.}
High-level semantic representations are important for understanding complex human interactions, while conventional engagement estimation methods mainly rely on low-level sensory features with limited contextual abstraction \cite{li2024dat,yu2023sliding}. Recent Multimodal Large Language Models (MLLMs) provide rich semantic representations that capture contextual information from multimodal interactions \cite{zhang2026affective,li2025emoverse,lin-etal-2024-video,zhang-etal-2023-video}. Recent studies further demonstrate their capability to analyze subtle affective preferences and high-level psychological information \cite{lian2025emoprefer,sabour-etal-2024-emobench}. Inspired by these advances, we leverage window-level MLLM representations to capture interaction context \cite{maaz-etal-2024-video} and integrate them with local temporal behaviors for context-aware engagement prediction.

% 可以适当的加一些卡通符号
\begin{figure*}[t]
    \centering
    \includegraphics[width=0.96\textwidth]{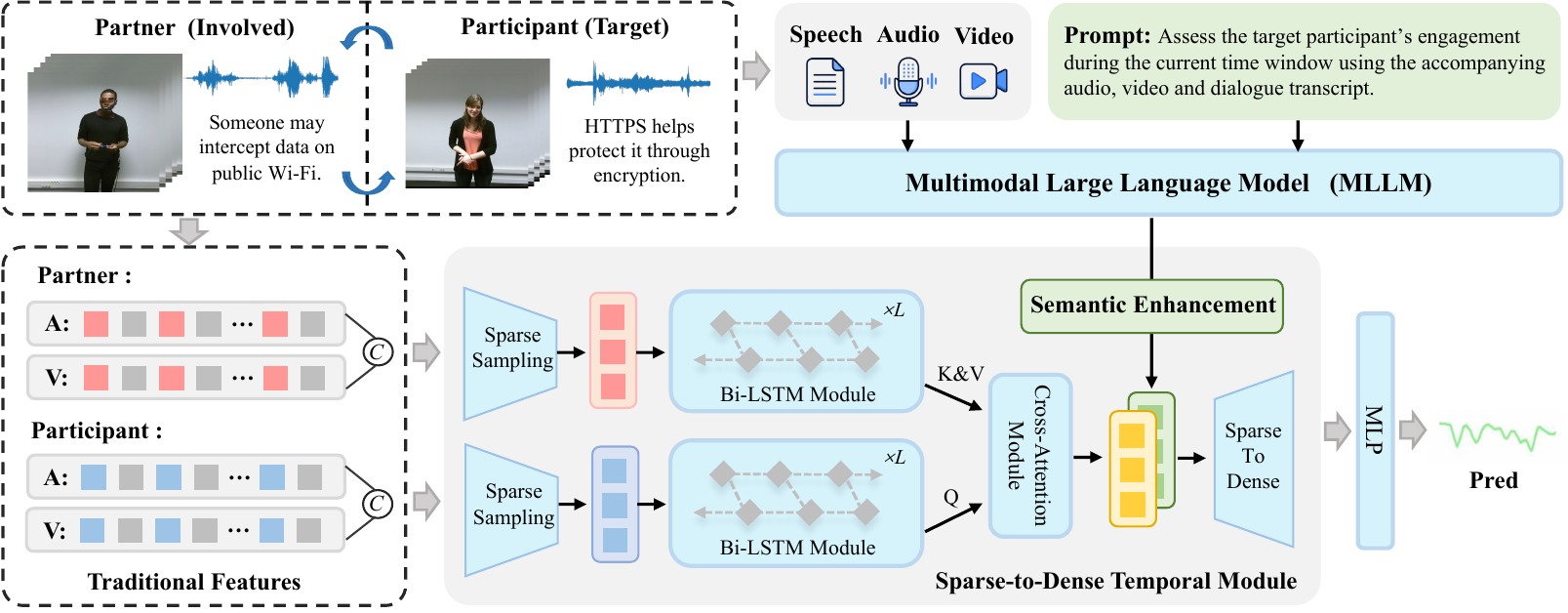}
    \caption{Overall framework of the proposed SEST model. Participant and partner audio-visual streams are encoded through a sparse temporal pathway, while MLLM-based semantic embeddings provide window-level contextual information. The semantic and temporal representations are fused and recovered to dense frame-level engagement predictions for the participant.}
    \label{fig:model_framework}
\end{figure*}
\section{Method}

% As illustrated in Figure~\ref{fig:model_framework}, the proposed SEST framework couples MLLM-based semantic enhancement modeling with a sparse-to-dense temporal module for multimodal dynamic engagement estimation. We first introduce the task formulation in Section~\ref{sec:task}. Section~\ref{sec:semantic} presents MLLM-based Semantic Enhancement Modeling (MSEM), which derives dialogue-aware semantic embeddings from interaction windows to complement frame-level behavioral features. Section~\ref{sec:sparse} then describes the Sparse-to-Dense Temporal Module (SDTM), where dense audio-visual sequences are transformed into compact temporal representations, enhanced through participant-partner interaction and semantic-temporal fusion, and recovered for continuous engagement prediction. Finally, Section~\ref{sec:objective} presents the dense prediction head and training objective.

As illustrated in Figure~\ref{fig:model_framework}, the proposed SEST framework couples MLLM-based Semantic Enhancement Modeling (MSEM) with a Sparse-to-Dense Temporal Module (SDTM) for multimodal dynamic engagement estimation. Section~\ref{sec:task} first presents the task formulation. Section~\ref{sec:semantic} then introduces MSEM, followed by the description of SDTM in Section~\ref{sec:sparse}. Finally, Section~\ref{sec:objective} presents the prediction head and training objective.

% \subsection{Task Formulation}
% \label{sec:task}

% Given a multimodal conversation corpus $\mathcal{D}$, we represent it as a set of samples $\{x_1,x_2,\ldots,x_{|\mathcal{D}|}\}$. Each sample $x$ comprises an audio-visual-text interaction segment, where the audio-visual sequence is denoted as $\mathbf{S}=\{\mathbf{S}_1,\mathbf{S}_2,\ldots,\mathbf{S}_N\}$ spanning $N$ frames. Each frame $\mathbf{S}_i=\{\mathbf{s}_i^a,\mathbf{s}_i^v\}$ contains synchronized audio and visual observations, where $a$ and $v$ denote the audio and visual modalities, respectively. In addition, the segment is associated with text modality $\mathbf{T}$, which represents the dialogue text or transcript within the same temporal window. The objective of the engagement estimation task is to learn a mapping $f:x\mapsto\mathbf{Y}$, where $x=\{\mathbf{S},\mathbf{T}\}$ denotes the multimodal input sample and $\mathbf{Y}=\{y_1,y_2,\ldots,y_N\}$ denotes the frame-level engagement score sequence. For each frame, $y_i\in[0,1]$ represents the continuous engagement score at the $i$-th frame. Accordingly, the model predicts a continuous engagement sequence $\hat{\mathbf{Y}}=\{\hat{y}_1,\hat{y}_2,\ldots,\hat{y}_N\}$ for the input sample $x$.

\subsection{Task Formulation}
\label{sec:task}

% Let $\mathcal{D}=\{x_1,x_2,\ldots,x_{|\mathcal{D}|}\}$ denote a multimodal conversation corpus, where each sample $x$ is an audio-visual-text interaction segment represented as $x=\{\mathbf{S},\mathbf{T}\}$. Here, $\mathbf{S}=\{\mathbf{S}_1,\mathbf{S}_2,\ldots,\mathbf{S}_N\}$ denotes the frame-level audio-visual sequence spanning $N$ frames, and each frame $\mathbf{S}_i=\{\mathbf{s}_i^a,\mathbf{s}_i^v\}$ contains synchronized audio and visual observations, where $a$ and $v$ indicate the audio and visual modalities, respectively. The text modality is denoted as $\mathbf{T}$, representing the dialogue text or transcript corresponding to the same interaction segment. Dynamic engagement estimation aims to learn a regression or classification function $f:x\mapsto\mathbf{Y}$, where $\mathbf{Y}=\{y_1,y_2,\ldots,y_N\}$ denotes the frame-level engagement annotation sequence. For continuous annotations, $y_i\in[0,1]$ denotes the engagement score at the $i$-th frame; for discrete annotations, $y_i\in\{1,2,\ldots,C\}$ denotes the engagement category among $C$ classes. Given the input sample $x$, the model outputs the predicted engagement sequence $\hat{\mathbf{Y}}=\{\hat{y}_1,\hat{y}_2,\ldots,\hat{y}_N\}$.

Let $\mathcal{D}=\{x_1,x_2,\ldots,x_{|\mathcal{D}|}\}$ denote a multimodal conversation corpus, where each sample $x$ is an audio-visual-text interaction segment represented as $x=\{\mathbf{S},\mathbf{T}\}$. Here, $\mathbf{S}=\{\mathbf{S}_1,\mathbf{S}_2,\ldots,\mathbf{S}_N\}$ denotes the frame-level audio-visual sequence spanning $N$ frames, and each frame $\mathbf{S}_i=\{\mathbf{s}_i^a,\mathbf{s}_i^v\}$ contains synchronized audio and visual observations, where $a$ and $v$ indicate the audio and visual modalities, respectively. The text modality is denoted as $\mathbf{T}$, representing the dialogue text or transcript corresponding to the same interaction segment. Dynamic engagement estimation aims to learn a mapping $f:x\mapsto\mathbf{Y}$, where $\mathbf{Y}=\{y_1,y_2,\ldots,y_N\}$ denotes the frame-level engagement annotation sequence. Depending on the annotation type, $y_i$ can be either a continuous engagement score or a discrete engagement category. Given the input sample $x$, the model outputs the predicted engagement sequence $\hat{\mathbf{Y}}=\{\hat{y}_1,\hat{y}_2,\ldots,\hat{y}_N\}$.

\subsection{MLLM-based Semantic Enhancement Modeling}
\label{sec:semantic}

We introduce MLLM-based Semantic Enhancement Modeling (MSEM) to enhance temporal engagement modeling with high-level conversational semantics. Specifically, MSEM leverages Multimodal Large Language Models (MLLMs) to extract window-level semantic embeddings from interaction segments, which provide dialogue-aware contextual information complementary to frame-level audio-visual representations.

% \subsubsection{MLLM-based Semantic Feature Extraction}

% Given an interaction window $\mathcal{W}$, we construct a multimodal input that contains the audio, visual, and textual information within the same temporal segment. Specifically, the window consists of the participant-side information and the partner-side information, denoted as $\mathcal{W}^{p}$ and $\mathcal{W}^{q}$, respectively, where $p$ indicates the participant whose engagement is to be estimated and $q$ indicates the interaction partner. A task-specific prompt $\mathcal{P}$ is designed to guide the MLLM to focus on engagement-related cues, such as the participant's behavioral state, the partner's response, and the conversational context.

% For each interaction window, the MLLM-based semantic encoder $\phi_{\mathrm{MLLM}}(\cdot)$ is used to extract semantic embeddings for the participant and the partner:
% \[
% \mathbf{e}^{p}=\phi_{\mathrm{MLLM}}(\mathcal{W}^{p},\mathcal{P}), 
% \quad
% \mathbf{e}^{q}=\phi_{\mathrm{MLLM}}(\mathcal{W}^{q},\mathcal{P}),
% \]
% where $\mathbf{e}^{p}$ and $\mathbf{e}^{q}$ denote the window-level semantic embeddings of the participant and the partner, respectively. Different from frame-level audio-visual features, these embeddings summarize high-level semantic and contextual information over the entire interaction window. Therefore, they provide complementary contextual cues for modeling engagement dynamics beyond local behavioral patterns.

\subsubsection{\bfseries MLLM-based Semantic Feature Extraction}

Given an interaction window $\mathcal{W}$, we construct multimodal inputs that contain audio, visual, and textual information. Since engagement is interaction-dependent, we organize the input into participant-side and partner-side representations, denoted as $\mathcal{W}^{p}$ and $\mathcal{W}^{q}$, respectively. Formally, each role-specific input is represented as
\[
\mathcal{W}^{r}=\{\mathbf{A}^{r},\mathbf{V}^{r},\mathbf{T}\}, \quad r\in\{p,q\},
\]
where $p$ indicates the participant whose engagement is to be estimated, $q$ indicates the interaction partner, $\mathbf{A}^{r}$ and $\mathbf{V}^{r}$ denote the audio and visual information of role $r$, and $\mathbf{T}$ denotes the multi-turn dialogue text related to the current interaction segment. A task-specific prompt $\mathcal{P}$ is designed to guide the MLLM to focus on engagement-related cues, including the participant's behavioral state, the partner's response, and the conversational context.

For each interaction window, we employ Gemini Embedding 2 \cite{shanbhogue2026gemini} via the Gemini API as a frozen semantic encoder $\phi_{\mathrm{MLLM}}(\cdot)$. The participant- and partner-side semantic representations are extracted as
\[
\mathbf{E}^{p}=\phi_{\mathrm{MLLM}}(\mathcal{W}^{p},\mathcal{P}), 
\quad
\mathbf{E}^{q}=\phi_{\mathrm{MLLM}}(\mathcal{W}^{q},\mathcal{P}),
\]
where $\mathbf{E}^{p}$ and $\mathbf{E}^{q}$ denote the window-level semantic representations of the participant and the partner, respectively. These representations summarize high-level semantic and contextual information over the interaction window, thereby providing complementary contextual cues for modeling engagement dynamics beyond local behavioral patterns.

\subsubsection{\bfseries Participant-Partner Semantic Relation Modeling}

To capture interaction-level semantic relations, we construct a semantic relation representation $\mathbf{R}^{s}$ from the MLLM-based semantic representations of the participant and the partner. Specifically, we concatenate the participant representation, the partner representation, their element-wise absolute difference, and their element-wise product:
\[
\mathbf{R}^{s}=
[
\mathbf{E}^{p};
\mathbf{E}^{q};
|\mathbf{E}^{p}-\mathbf{E}^{q}|;
\mathbf{E}^{p}\odot\mathbf{E}^{q}
],
\]
where the superscript $s$ denotes semantic representation, $[\cdot;\cdot]$ denotes feature concatenation, $|\cdot|$ denotes the element-wise absolute difference, and $\odot$ denotes the element-wise product. This representation preserves individual semantic states while encoding their discrepancy and interaction consistency.

The semantic relation representation $\mathbf{R}^{s}$ is then projected into a task-specific semantic space:
\[
\mathbf{Z}^{s}=\mathrm{MLP}_{s}(\mathbf{R}^{s}),
\]
where $\mathrm{MLP}_{s}$ denotes a lightweight projection network. The resulting representation $\mathbf{Z}^{s}$ serves as the window-level semantic feature and is fused with temporal representations in the subsequent sparse-to-dense temporal module.

\subsection{Sparse-to-Dense Temporal Module}
\label{sec:sparse}

The Sparse-to-Dense Temporal Module (SDTM) aims to model engagement dynamics efficiently while preserving dense frame-level prediction. It transforms audio-visual features into a sparse temporal space, integrates interaction and semantic information, and finally recovers continuous engagement predictions at the original frame resolution.

\subsubsection{\bfseries Multimodal Feature Extraction}

% For each interaction window, we extract a diverse set of multimodal features to capture the multi-faceted behavioral dynamics of both the target participant and the partner. From the acoustic modality, we combine low-level acoustic descriptors from eGeMAPS \cite{eyben2010opensmile} with high-level speech representations from a pretrained Whisper \cite{radford2023robust} model, ensuring a balance between prosodic nuances and semantic-rich speech context. Regarding the visual modality, we integrate global scene context extracted via CLIP \cite{radford2021learning} and Swin Transformer \cite{liu2021Swin} with fine-grained behavioral cues, including facial landmarks from OpenFace \cite{baltruvsaitis2016openface} and body keypoints from OpenPose \cite{cao2017realtime}. These modality-specific features are projected into a unified latent space and concatenated to form frame-level feature sequences denoted as $\mathbf{X}^{p}\in\mathbb{R}^{N\times D}$ and $\mathbf{X}^{q}\in\mathbb{R}^{N\times D}$ for the participant and the partner, respectively.

For each interaction window, we extract multimodal features from the audio-visual streams of both the participant and the partner. For the acoustic modality, we use eGeMAPS~\cite{eyben2010opensmile} and Whisper~\cite{radford2023robust} features to capture low-level acoustic patterns and high-level speech representations. For the visual modality, we combine global visual representations from CLIP~\cite{radford2021learning} and Swin Transformer~\cite{liu2021Swin} with fine-grained behavioral cues from OpenFace~\cite{baltruvsaitis2016openface} and OpenPose~\cite{cao2017realtime}. These modality-specific features are projected into a unified latent space and concatenated into frame-level feature sequences, denoted as $\mathbf{X}^{p}\in\mathbb{R}^{N\times D}$ and $\mathbf{X}^{q}\in\mathbb{R}^{N\times D}$ for the participant and the partner, respectively.

\subsubsection{\bfseries Sparse-to-Dense Engagement Prediction}

To perform temporal modeling in a compact sequence space, SDTM first sparsifies the participant and partner feature sequences by interval sampling. Specifically, given $\mathbf{X}^{r}=\{\mathbf{x}^{r}_{1},\mathbf{x}^{r}_{2},\ldots,\mathbf{x}^{r}_{N}\}$, $r\in\{p,q\}$, we sample one frame every $\lambda$ frames and obtain the sparse sequence $\mathbf{H}^{r}_{s}=\{\mathbf{x}^{r}_{1},\mathbf{x}^{r}_{1+\lambda},\ldots,\mathbf{x}^{r}_{1+(M-1)\lambda}\}\in\mathbb{R}^{M\times D}$, where $\lambda$ is the sampling interval and $M=\lceil N/\lambda\rceil$ denotes the sparse sequence length.

The sparse sequences are then fed into an $L$-layer bidirectional LSTM for temporal encoding. By modeling dependencies in both forward and backward directions, the encoder extracts compact temporal dynamics from the reduced sequence space:
\[
\mathbf{Z}^{r}_{t}=\mathrm{BiLSTM}^{L}_{r}(\mathbf{H}^{r}_{s}),
\quad r\in\{p,q\},
\]
where $\mathbf{Z}^{p}_{t}$ and $\mathbf{Z}^{q}_{t}$ represent the encoded sparse temporal features of the participant and the partner, respectively. To model cross-role interaction, we apply cross-attention from the participant stream to the partner stream. The query is obtained from the participant representation as $\mathbf{Q}=\mathbf{Z}^{p}_{t}\mathbf{W}_{Q}$, while the key and value are obtained from the partner representation as $\mathbf{K}=\mathbf{Z}^{q}_{t}\mathbf{W}_{K}$ and $\mathbf{V}=\mathbf{Z}^{q}_{t}\mathbf{W}_{V}$, where $\mathbf{W}_{Q}$, $\mathbf{W}_{K}$, and $\mathbf{W}_{V}$ are learnable projection matrices. The interaction-aware temporal representation is computed as
\[
\mathbf{Z}_{c}=
\mathrm{softmax}\left(\frac{\mathbf{Q}\mathbf{K}^{\top}}{\sqrt{d_{k}}}\right)\mathbf{V},
\]
where $d_{k}$ is the key dimension, and $\mathbf{Z}_{c}$ denotes the interaction-aware sparse temporal representation.

The MLLM-based semantic representation $\mathbf{Z}^{s}$ is then repeated along the temporal dimension and concatenated with the interaction-aware feature:
\[
\mathbf{Z}_{f}
=
\mathrm{Concat}
\left(
\mathbf{Z}_{c},
\mathrm{Repeat}(\mathbf{Z}^{s},M)
\right),
\]
where $\mathrm{Repeat}(\cdot,M)$ aligns the window-level semantic representation with the sparse sequence length and $\mathbf{Z}_{f}$ denotes the semantic-enhanced sparse temporal representation. Finally, the fused sparse representation is recovered to the original temporal resolution through a transposed one-dimensional convolution:
\[
\mathbf{H}_{d}
=
\mathrm{ConvTrans1D}(\mathbf{Z}_{f}),
\]
\[
\hat{\mathbf{Y}}
=
\mathrm{MLP}(\mathbf{H}_{d}),
\]
where $\mathrm{ConvTrans1D}(\cdot)$ denotes the transposed one-dimensional convolution for sparse-to-dense recovery, $\mathbf{H}_{d}\in\mathbb{R}^{N\times D_h}$ is the recovered dense temporal representation, and $\hat{\mathbf{Y}}=\{\hat{y}_{1},\hat{y}_{2},\ldots,\hat{y}_{N}\}$ is the predicted frame-level engagement sequence. Through this process, SDTM performs compact temporal modeling while producing predictions at the original annotation resolution.

\subsection{Training Objective}
\label{sec:objective}

% For continuous engagement annotations, we optimize the model using the Concordance Correlation Coefficient (CCC) loss. Given the predicted sequence $\hat{\mathbf{Y}}$ and the ground-truth sequence $\mathbf{Y}$, CCC is defined as
% \[
% \rho_{c} =
% \frac{2\rho_{\hat{\mathbf{Y}},\mathbf{Y}}\sigma_{\hat{\mathbf{Y}}}\sigma_{\mathbf{Y}}}
% {\sigma_{\hat{\mathbf{Y}}}^{2}+\sigma_{\mathbf{Y}}^{2}+(\mu_{\hat{\mathbf{Y}}}-\mu_{\mathbf{Y}})^{2}},
% \]
% where $\rho_{\hat{\mathbf{Y}},\mathbf{Y}}$ denotes the Pearson correlation coefficient, and $\mu$ and $\sigma$ denote the mean and standard deviation, respectively. The CCC loss is formulated as
% \[
% \mathcal{L}_{\mathrm{CCC}}=1-\rho_{c}.
% \]

% For discrete engagement annotations, we use a differentiable Kappa loss to optimize agreement between predicted and ground-truth categories:
% \[
% \mathcal{L}_{\kappa}=1-\kappa,
% \]
% where $\kappa$ denotes the differentiable approximation of Cohen's Kappa. Accordingly, the overall objective is selected according to the annotation type:
% \[
% \mathcal{L}=
% \begin{cases}
% \mathcal{L}_{\mathrm{CCC}}, & \text{for continuous annotations},\\
% \mathcal{L}_{\kappa}, & \text{for discrete annotations}.
% \end{cases}
% \]

For continuous engagement annotations, we optimize the model using the Concordance Correlation Coefficient (CCC) loss. Given the predicted sequence $\hat{\mathbf{Y}}$ and the ground-truth sequence $\mathbf{Y}$, CCC is defined as
\[
\rho_{c} =
\frac{2\rho_{\hat{\mathbf{Y}},\mathbf{Y}}\sigma_{\hat{\mathbf{Y}}}\sigma_{\mathbf{Y}}}
{\sigma_{\hat{\mathbf{Y}}}^{2}+\sigma_{\mathbf{Y}}^{2}+(\mu_{\hat{\mathbf{Y}}}-\mu_{\mathbf{Y}})^{2}},
\]
where $\rho_{\hat{\mathbf{Y}},\mathbf{Y}}$ denotes the Pearson correlation coefficient, and $\mu$ and $\sigma$ denote the mean and standard deviation, respectively. The corresponding loss is $\mathcal{L}_{\mathrm{CCC}}=1-\rho_c$. For discrete engagement annotations, we use a differentiable Kappa loss, denoted as $\mathcal{L}_{\kappa}=1-\kappa$, where $\kappa$ is a differentiable approximation of Cohen's Kappa. The training objective is selected according to the annotation type:
\[
\mathcal{L}=
\begin{cases}
\mathcal{L}_{\mathrm{CCC}}, & \text{for continuous annotations},\\
\mathcal{L}_{\kappa}, & \text{for discrete annotations}.
\end{cases}
\]

\section{Experiments}
\subsection{Datasets}
 The challenge comprises five datasets evaluated using two metrics: the Concordance Correlation Coefficient (CCC) for the four conversational sets, and Cohen’s Kappa (CK) for the PInSoRo classification task. The final ranking is determined by the aggregate performance across all domains.

{\bfseries NoXi Series datasets}. The challenge includes three subsets from the NoXi corpus: {\bfseries NoXi Base \cite{cafaro2017noxi}} provides 48 training and 16 test sessions of expert-novice interactions in English, French, and German; {\bfseries NoXi-Additional \cite{cafaro2017noxi,muller2024multimediate}} contains 12 test sessions in Arabic, Italian, Indonesian, and Spanish to evaluate cross-lingual generalization; and {\bfseries NoXi-J \cite{funk2024multilingual}} comprises 10 test sessions in Japanese and Mandarin Chinese, extending coverage to East Asian cultural contexts. All sessions include synchronized audio-visual streams at 25 fps with frame-level engagement labels in [0, 1]. Both raw videos and pre-extracted features are utilized across these subsets.

{\bfseries MPIIGroupInteraction (MPIIGI) dataset \cite{muller2018detecting,muller2021multimediate}} includes 6 test sessions, and each 20-minute session involves four participants seated around a table discussing a controversial topic in German. MPIIGI features multi-party seated discussions, with feature sampling rates aligned with the NoXi datasets.

{\bfseries PInSoRo \cite{lemaignan2018pinsoro} dataset} captures child–child and child–robot play interactions among children aged 4–8. It comprises 56 sessions. Distinct from the continuous scales in other datasets, PInSoRo provides categorical annotations for task and social engagement. 

% \begin{table*}[!t]
% \centering
% \caption{Comparison of final results on the challenge datasets. CCC scores for NoXi Base, NoXi-J, and MPIIG are reported on the validation sets, while Test CCC Global and Test CK PInSoRo are reported on the test sets.}
% \label{tab:final_results}
% \begin{tabular}{ll ccccc}
% \toprule
% \textbf{Model} & \textbf{Method} & \textbf{Val CCC} & \textbf{Val CCC} & \textbf{Val CCC} & \textbf{Test CCC} & \textbf{Test CK} \\
%  &  & \textbf{NoXi Base} & \textbf{NoXi-J} & \textbf{MPIIG} & \textbf{Global} & \textbf{PInSoRo} \\
% \midrule
% Müller et al \cite{withanage25_mm}. & 25'Baseline & -  & -  & -  & 0.405  & -  \\
% Müller et al & 26'Baseline & -  & -  & -  & 0.450  & 0.160  \\
% Yu et al \cite{yu2023sliding}. & Sliding Window Seq2seq & 0.715  & 0.409 &  -  &  0.520 & - \\
% Li et al\cite{li2024dat}. & DAT, MGF & 0.846  & - & 0.504  &  -  & -  \\
% Yu et al\cite{ijcai2024p353}. & Dialogue Cross-Enhanced CEAM & 0.706 & 0.648 & -  & 0.562 & -  \\
% Zou et al. & BiLSTM+KAN+8$\times$SW & 0.831  & 0.657   & - & 0.678 & - \\
% Yu  et al\cite{yu2025generalizable}.  &DAPA   & 0.851   & 0.717  & 0.655  &0.700   & -  \\
% \textbf{Ours} & \textbf{SEST} & \textbf{0.864} & \textbf{0.738} & \textbf{0.687} & \textbf{0.708} & \textbf{0.520} \\
% \bottomrule
% \end{tabular}
% \end{table*}

\begin{table*}[!t]
\centering
\renewcommand{\arraystretch}{1.1}
\captionsetup{font=normalsize}
\normalsize
\caption{Comparison of final results on the challenge datasets. CCC scores for NoXi Base, NoXi-J, and MPIIGI are reported on the validation sets, while Test CCC Global, Test CK PInSoRo and Test Global are reported on the test sets.}
\label{tab:final_results}
\begin{tabular}{llcccccc}
\specialrule{1.0pt}{0pt}{2pt}
\textbf{Model} & \textbf{Method} 
& \textbf{Val CCC} & \textbf{Val CCC} & \textbf{Val CCC} 
& \textbf{Test CCC} & \textbf{Test CK} & \textbf{Test} \\
 & 
& \textbf{NoXi Base} & \textbf{NoXi-J} & \textbf{MPIIGI} 
& \textbf{Global} & \textbf{PInSoRo} & \textbf{Global} \\
\midrule
Müller et al.~\cite{withanage25_mm} & 25' Baseline & -- & -- & -- & 0.405 & --  & -- \\
Müller et al. & 26' Baseline & -- & -- & -- & 0.450 & 0.160  & 0.355 \\
Yu et al.~\cite{yu2023sliding} & Sliding Window Seq2seq & 0.715 & 0.409 & -- & 0.520 & --  & -- \\
Li et al.~\cite{li2024dat} & DAT, MGF & 0.846 & -- & 0.504 & -- & --  & --  \\
Yu et al.~\cite{ijcai2024p353} & Dialogue Cross-Enhanced CEAM & 0.706 & 0.648 & -- & 0.562 & --  & -- \\
Zou et al.~\cite{10.1145/3746027.3762078} & BiLSTM+KAN+8$\times$SW & 0.831 & 0.657 & -- & 0.678 & --  & --  \\
Yu et al.~\cite{yu2025generalizable} & DAPA & 0.851 & 0.717 & 0.655 & 0.700 & --  & -- \\
\rowcolor{lightblue}
\textbf{Ours} & \textbf{SEST} 
& \textbf{0.864} & \textbf{0.738} & \textbf{0.687} 
& \textbf{0.708} & \textbf{0.520}  & \textbf{0.65} \\
\specialrule{1.0pt}{2pt}{0pt}
\end{tabular}
\end{table*}

\subsection{Experimental Setup}
We use synchronized multimodal features, including eGeMAPS (88D), Whisper (1280D), CLIP (512D), Swin Transformer (768D), OpenPose (139D), and OpenFace 3.0 (21D). All features are projected to 512 dimensions, concatenated, and normalized at the session level. We use 288-frame windows with a stride of 96 and dataset-specific sparse intervals ($\lambda=3$ for NoXi-Base, MPIIGI, and PInSoRo; $\lambda=4$ for NoXi-J). The sparse sequences are encoded by a 4-layer BiLSTM with a hidden dimension of 256. The 3072-D MLLM embeddings are projected and fused with temporal features for dense prediction. Models are trained for 40 epochs using Adam with a batch size of 32 and an initial learning rate of $5\times10^{-5}$. We train five models with seeds 40--44 using exponential moving average (EMA) and ensemble their predictions. Overlapping-window predictions are averaged during inference.

% \begin{table}[!t]
% \centering
% \caption{CCC-Average denotes the average performance on NoXi Base, NoXi-J, NoXi-Additional, and MPIIG, while CK-Average denotes the performance on PInSoRo. Global is the overall final score.}
% \label{tab:leaderboard}
% \begin{tabular}{lccc}
% \toprule
% \textbf{Team} & \textbf{CCC-Average} & \textbf{CK-Average} & \textbf{Global} \\
% \midrule
% \textbf{HFUT-LMC(Ours)}~\first        & 0.71 & \bfseries 0.52 & \bfseries 0.65 \\
% \textbf{NUDT\_LDM-AF}~\second   & 0.73 & 0.46 & 0.64 \\
% \textbf{mltac}~\third           & 0.70 & 0.36 & 0.59 \\
% USTC\_IAT\_United               & \textbf{0.74} & 0.22 & 0.57 \\
% SZY\_MM                         & 0.65 & 0.34 & 0.55 \\
% RGroupEngaged                   & 0.68 & 0.28 & 0.55 \\
% mvechat                         & 0.58 & 0.18 & 0.45 \\
% 26'Baseline                     & 0.45 & 0.16 & 0.36 \\
% \bottomrule
% \end{tabular}
% \end{table}

\begin{table}[!t]
\centering
\renewcommand{\arraystretch}{1.1}
\captionsetup{font=normalsize}
\normalsize
\setlength{\tabcolsep}{3.5pt}
\caption{Final team rankings on the challenge leaderboard. CCC-Avg. denotes the average performance on NoXi Base, NoXi-J, NoXi-Additional, and MPIIGI, while CK-Avg. denotes the performance on PInSoRo. Global is the overall final score.}
\label{tab:leaderboard}
\begin{tabular}{@{}lccc@{}}
\specialrule{1.0pt}{0pt}{2pt}
\textbf{Team} & \textbf{CCC-Avg.} & \textbf{CK-Avg.} & \textbf{Global} \\
\midrule
\rowcolor{lightblue}
\textbf{HFUT-LMC (Ours)}~\first & 0.71 & \textbf{0.52} & \textbf{0.65} \\
\textbf{NUDT\_LDM-AF}~\second & 0.73 & 0.46 & 0.64 \\
\textbf{mltac}~\third & 0.70 & 0.36 & 0.59 \\
USTC\_IAT\_United & \textbf{0.74} & 0.22 & 0.57 \\
SZY\_MM & 0.65 & 0.34 & 0.55 \\
RGroupEngaged & 0.68 & 0.28 & 0.55 \\
mvechat & 0.58 & 0.18 & 0.45 \\
26' Baseline & 0.45 & 0.16 & 0.36 \\
\specialrule{1.0pt}{2pt}{0pt}
\end{tabular}
\end{table}

% \subsection{Final Results}

% Table 1 compares our SEST model with recent state-of-the-art frameworks and challenge baselines. Our method achieves superior performance across all evaluated subsets, reaching a Global CCC of 0.710. Notably, SEST significantly outperforms previous models on the complex NoXi-J (0.741) and MPIIG (0.680) datasets. This suggests that the integration of window-level MLLM semantic context effectively complements traditional behavioral features, especially in cross-cultural and multi-party interaction scenarios where overt cues may be ambiguous. Furthermore, SEST demonstrates strong generalization on the PInSoRo test set (CK: 0.520), confirming its robustness across diverse interaction paradigms and age groups.

% Table 2 presents the final team rankings for the MEE2026 Challenge. Our team (HFUT-LMC) secured the first place with an overall Global score of 0.65. While several teams achieved competitive results on the NoXi-based CCC-average metric (e.g., USTC\_IAT\_United at 0.74), our model maintains the highest performance on the PInSoRo dataset (CK-average:0.52). This balanced excellence across conversational human-human interaction and child-robot play-driven dynamics underscores the effectiveness of our semantic-enhanced sparse temporal modeling in capturing universal engagement patterns.

\subsection{Final Results}

Table~\ref{tab:final_results} summarizes the comparison between SEST and recent methods on the evaluated challenge datasets. SEST achieves the best results on the reported validation subsets, with CCC scores of 0.864 on NoXi Base, 0.738 on NoXi-J, and 0.687 on MPIIGI. On the test evaluation, SEST obtains a Test CCC Global score of 0.708 and a Test CK score of 0.520, showing competitive performance under both continuous and discrete engagement evaluation settings. Table~\ref{tab:leaderboard} further reports the final MEE2026 Challenge leaderboard. Based on the combined CCC-Average and CK-Average scores, our team, HFUT-LMC, ranks first with a final Global score of 0.65.

\subsection{Ablation Studies}

% \subsubsection{\bfseries Effectiveness of SDTM and MSEM Modules}
% We first evaluate the individual contributions of the  Sparse-to-Dense Temporal Module(SDTM) and MLLM-based Semantic Enhancement Modeling(MSEM). The baseline model consists of a standard BiLSTM operating on dense multimodal features. As shown in Table 3, the integration of the MSEM module leads to a consistent performance gain across all datasets. This improvement demonstrates that by sparsely sampling redundant frame-level features and recovering dense representations, the model can effectively filter out local temporal noise and focus on long-term engagement trends. Furthermore, the addition of the MSEM module, which provides window-level semantic context via MLLM embeddings, significantly boosts performance on the NoXi-Base and NoXi-J subsets. Notably, for the PInSoRo dataset, the MSEM module is omitted due to the absence of raw video data, yet the SD module still maintains its superiority over the dense baseline.

\subsubsection{\bfseries Effectiveness of SDTM and MSEM Modules}

We first evaluate the individual contributions of the Sparse-to-Dense Temporal Module (SDTM) and MLLM-based Semantic Enhancement Modeling (MSEM). The baseline is a standard BiLSTM built on dense multimodal features. As shown in Table~\ref{tab:ablation_overall}, SDTM consistently improves the baseline, with a clear gain on NoXi-J, reaching 0.726 compared with 0.693. This shows that sparse temporal modeling provides a compact yet effective representation for dense engagement prediction. MSEM also improves the baseline by introducing window-level semantic context, achieving 0.706 on NoXi-J when used alone. Combining SDTM and MSEM yields the best results on the continuous engagement subsets, with 0.864 on NoXi-Base, 0.738 on NoXi-J, and 0.687 on MPIIGI. For PInSoRo, MSEM is not used due to the absence of raw video data, while SDTM improves the validation CK from 0.464 to 0.533.
% For PInSoRo, MSEM is not used due to the absence of raw video data, while SDTM still improves the baseline, reaching 0.533.

% \begin{table}[t]
% \centering
% \caption{Overall ablation study of the proposed SEST components. The best results are highlighted in bold.}
% \label{tab:ablation_overall}
% \begin{tabular}{cc cccc}
% \toprule
% \textbf{SDTM} & \textbf{MSEM} & \textbf{NoXi-Base} & \textbf{NoXi-J} & \textbf{MPIIG} & \textbf{PInSoRo} \\ 
% \midrule
%  &  & 0.840 & 0.693 & 0.647 & 0.464 \\
% \cmark &  & 0.853 & 0.726 & 0.670 & \textbf{0.533} \\
%  & \cmark & 0.845 & 0.706 & 0.653 & -- \\
% \cmark & \cmark & \textbf{0.864} & \textbf{0.738} & \textbf{0.687} & -- \\ 
% \bottomrule
% \end{tabular}
% \end{table}

\begin{table}[!t]
\centering
\renewcommand{\arraystretch}{1.1}
\captionsetup{font=normalsize}
\normalsize
\setlength{\tabcolsep}{3pt}
\caption{Overall ablation results on the validation sets. NoXi-Base, NoXi-J, and MPIIGI are evaluated by CCC, while PInSoRo is evaluated by Cohen's Kappa. The best results are highlighted in bold.}
\label{tab:ablation_overall}
\begin{tabular}{@{}cccccc@{}}
\specialrule{1.0pt}{0pt}{2pt}
\textbf{SDTM} & \textbf{MSEM} & \textbf{NoXi-Base} & \textbf{NoXi-J} & \textbf{MPIIGI} & \textbf{PInSoRo} \\ 
 &  & \textbf{CCC} & \textbf{CCC} & \textbf{CCC} & \textbf{CK} \\
\midrule
 &  & 0.840 & 0.693 & 0.647 & 0.464 \\
\cmark &  & 0.853 & 0.726 & 0.670 & \textbf{0.533} \\
 & \cmark & 0.845 & 0.706 & 0.653 & -- \\
\rowcolor{lightblue}
\cmark & \cmark & \textbf{0.864} & \textbf{0.738} & \textbf{0.687} & -- \\ 
\specialrule{1.0pt}{2pt}{0pt}
\end{tabular}
\end{table}

% \subsubsection{\bfseries Sensitivity to Sparse Sampling Factor}
% Figure 2 shows the effect of different sparse factors on engagement estimation performance. Across NoXi-Base, NoXi-J, and MPIIGI, performance consistently improves from dense input to moderate sparse sampling, and then drops when the sparse factor becomes too large. The best results are achieved at factor 3 for NoXi-Base and MPIIGI, and factor 4 for NoXi-J. This trend indicates that dense frame-level modeling contains temporal redundancy, while moderate sparse sampling provides a better balance between efficiency and information preservation. However, overly aggressive sparsification may remove useful local temporal cues and lead to performance degradation.

\subsubsection{\bfseries Impact of MLLM Integration Strategies}

We further evaluate different MLLM integration strategies. As shown in Table~\ref{tab:ablation_fusion}, SEST adopts Late Fusion, where the MLLM-based semantic representation is fused after temporal modeling. We compare it with MLLM Embeddings Only and Early Fusion, where the MLLM embedding is directly concatenated with conventional features before temporal encoding. Late Fusion achieves the best results, with CCC scores of 0.864 on NoXi-Base and 0.738 on NoXi-J. In contrast, MLLM Embeddings Only performs much lower, indicating that semantic embeddings alone are insufficient for frame-level prediction. Early Fusion also underperforms Late Fusion, suggesting that window-level semantic information is more effective when used as complementary context after temporal representation learning.

% \begin{table}[t]
% \centering
% \caption{Comparison of MLLM integration strategies (CCC Score).}
% \label{tab:ablation_fusion}
% \begin{tabular}{lcc}
% \toprule
% \textbf{Strategy} & \textbf{NoXi-Base} & \textbf{NoXi-J} \\ 
% \midrule
% MLLM Embeddings Only        & 0.649 & 0.551 \\
% Early Fusion                & 0.8536 & 0.674 \\
% \midrule
% \textbf{Late Fusion (Ours)} & \textbf{0.864} & \textbf{0.738} \\ 
% \bottomrule
% \end{tabular}
% \end{table}

\begin{table}[t]
\centering
\renewcommand{\arraystretch}{1.1}
\captionsetup{font=normalsize}
\normalsize
\setlength{\tabcolsep}{4pt}
\caption{Comparison of MLLM integration strategies in terms of CCC score.}
\label{tab:ablation_fusion}
\begin{tabular}{@{}lcc@{}}
\specialrule{1.0pt}{0pt}{2pt}
\textbf{Strategy} & \textbf{NoXi-Base} & \textbf{NoXi-J} \\ 
\midrule
MLLM Embeddings Only & 0.649 & 0.551 \\
Early Fusion & 0.854 & 0.674 \\
\midrule
\rowcolor{lightblue}
\textbf{Late Fusion (Ours)} & \textbf{0.864} & \textbf{0.738} \\ 
\specialrule{1.0pt}{2pt}{0pt}
\end{tabular}
\end{table}

\subsubsection{\bfseries Sensitivity to Sparse Sampling Interval}

\begin{figure}[t]
    \centering
    \includegraphics[width=0.98\linewidth]{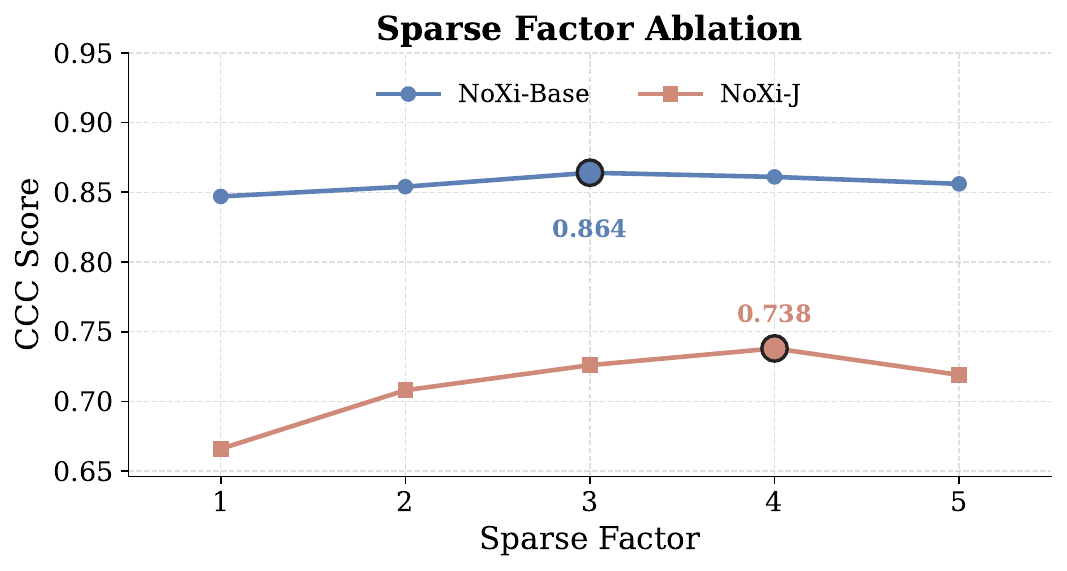}
    \caption{Dataset-specific effect of the sparse factor on CCC performance. The stars indicate the optimal factor for each dataset. }
    \label{fig:sparse_factor_ablation}
\end{figure}

Figure~\ref{fig:sparse_factor_ablation} shows the effect of the sparse sampling interval $\lambda$ on engagement estimation performance. Across NoXi-Base and NoXi-J, the performance improves when moving from dense input to moderate sparse sampling, but decreases when $\lambda$ becomes too large. The best results are obtained at $\lambda=3$ for NoXi-Base and $\lambda=4$ for NoXi-J. This suggests that moderate sparse sampling reduces temporal redundancy while preserving sufficient behavioral cues, whereas overly large $\lambda$ may discard useful local information and degrade performance.

% \subsubsection{\bfseries Impact of MLLM Integration Strategies}
% Finally, we explore different strategies for integrating window-level MLLM embeddings into the temporal branch. As shown in Table 4, We compare three configurations: (1) MLLM Only, which predicts engagement solely from semantic embeddings; (2) Early Fusion, where embeddings are repeated and concatenated with sparse features before the BiLSTM encoder; and (3) Late Fusion (SEST), where semantic context is fused with the recovered dense temporal hidden states. Results in Table 4 show that Late Fusion achieves the highest accuracy. While MLLM embeddings provide strong global context, they lack the temporal resolution to track rapid engagement shifts independently. Early fusion performs sub-optimally as it tends to dilute the dynamic behavioral signals with static window-level information. Our late fusion strategy effectively treats the MLLM embedding as a high-level semantic prior, successfully guiding the continuous prediction of the temporal branch.

\subsection{Analysis}

% \subsubsection{\bfseries Impact of Window Size and Temporal Sparsity}
% We further investigate the relationship between temporal window length and prediction accuracy. As shown in the line chart (Fig. [Y]), without sparse modeling, the CCC score initially increases with window size (from 32 to 128 frames) as the model gains more temporal context. However, beyond a certain threshold (e.g., 288 frames), the performance begins to degrade. We attribute this to the fact that for a fixed parameter budget, dense modeling of excessively long windows leads to increased computational redundancy and optimization difficulty, causing the model to overfit to local noise rather than global trends.

% Upon introducing our Sparse Temporal Modeling, a distinct shift in the optimal window size is observed. The SEST framework achieves its peak performance at a larger window size (e.g., 288 frames) compared to the dense baseline. This is because sparse sampling effectively filters out frame-level redundancy and reduces the receptive field's noise, allowing the BiLSTM encoder to aggregate information over longer durations without a proportional increase in complexity. This results in a "Sparse-to-Dense" benefit: the model maintains the fine-grained resolution required for continuous prediction while effectively modeling long-range interaction dependencies.

\subsubsection{\bfseries Impact of Sparse Sampling on Optimal Window Size}

We further analyze how sparse sampling affects the optimal temporal window size. As shown in Figure 3, the dense baseline achieves its best performance at a moderate window length of 160 frames, while further increasing the window size does not bring additional improvement. This may be caused by the increased redundancy and local noise introduced by dense modeling over longer windows, which can make temporal learning less effective.

After introducing sparse sampling, the optimal window shifts to a longer temporal range. The sparse model achieves its best performance at 288 frames, reaching a CCC score of 0.726. This indicates that sparse sampling reduces redundant frame-level observations and allows the model to exploit broader temporal context more effectively. In this sense, the sparse-to-dense design helps extend the effective temporal window while still recovering dense frame-level predictions.

\begin{figure}[t]
    \centering
    \includegraphics[width=0.98\linewidth]{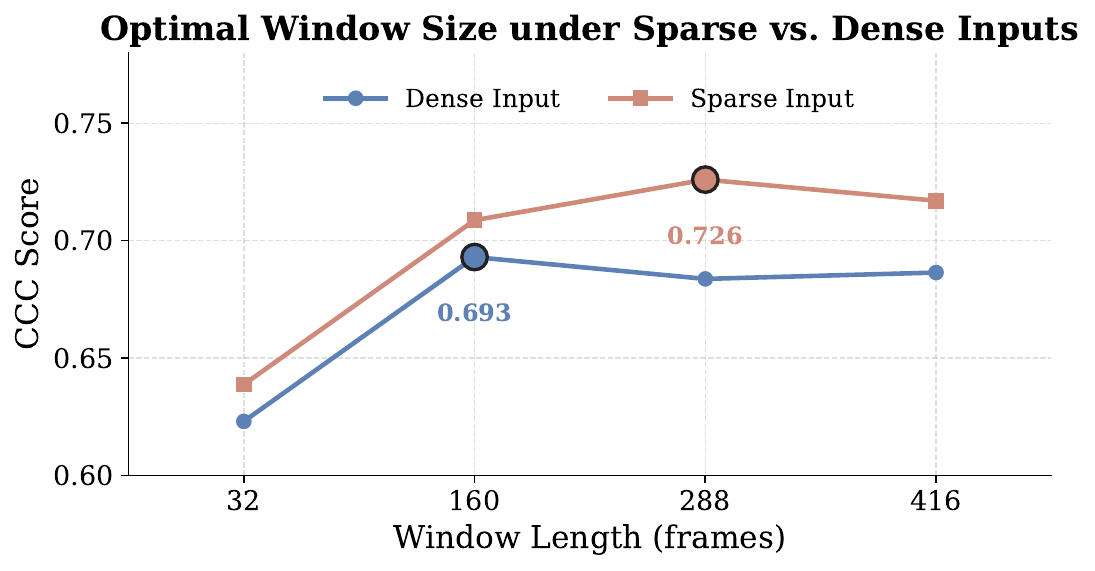}
    \caption{Comparison of CCC performance under different temporal window sizes on the NoXi-J validation set. Dense input and sparse input are evaluated to analyze the effect of sparse sampling on window-size selection.}
    \label{fig:window_length_ablation}
\end{figure} 

\subsubsection{\bfseries Visualization of Engagement Trajectories}Figure 4 visualizes the engagement predictions of our SEST model against the ground truth for representative validation sessions from NoXi-Base (top row) and NoXi-J (bottom row). The two subsets exhibit noticeably different annotation patterns: NoXi-Base contains more dynamic and continuous fluctuations, whereas NoXi-J shows a step-like pattern with sustained levels and abrupt transitions. Despite these differences, our model consistently tracks the ground-truth curves in both settings, indicating good robustness and generalization across diverse temporal dynamics and annotation styles.

\begin{figure}[t]
    \centering
    \includegraphics[width=\linewidth]{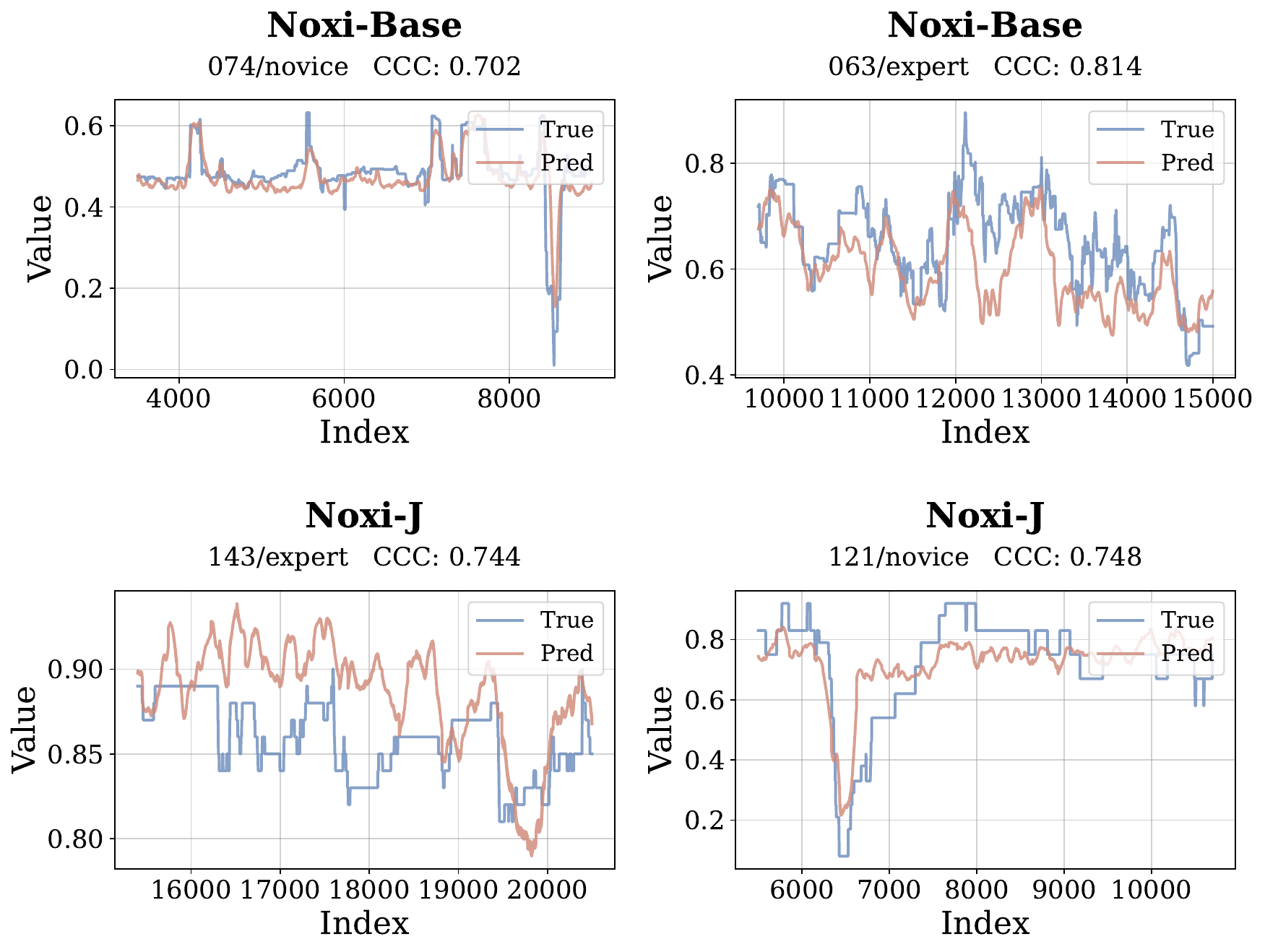}
    \caption{Visualization of engagement predictions on NoXi-Base (top) and NoXi-J (bottom). Blue lines denote ground truth; orange lines represent SEST predictions. }
    \label{fig:qualitative_results}
\end{figure}

% \section{Conclusion}
% In this paper, we proposed semantic-enhanced sparse temporal modeling for dynamic engagement estimation in multimodal interactions. The method addresses two practical limitations of existing dense multimodal modeling: insufficient window-level semantic context and redundant frame-level temporal input. By introducing large-model-based semantic enhancement and sparse-to-dense temporal modeling, the proposed framework combines global semantic context with compact temporal dynamics for dense engagement prediction. Experimental results on MEE2026 validate the effectiveness of the proposed design.

\section{Conclusion}

In this paper, we proposed SEST, a semantic-enhanced sparse temporal framework for dynamic engagement estimation in multimodal interactions. SEST revisits dense engagement prediction from two complementary aspects: semantic enrichment and temporal efficiency. On the semantic side, it leverages MLLM-based representations to incorporate high-level conversational context that is difficult to align with frame-level annotations. On the temporal side, it introduces a sparse-to-dense modeling strategy to capture broader engagement dynamics while maintaining dense prediction resolution. Experiments on the MEE2026 benchmark show that SEST achieves strong overall performance, and ablation studies verify the effectiveness of both semantic enhancement and sparse temporal modeling. These results highlight the importance of high-level conversational semantics and efficient sparse temporal modeling for dynamic engagement estimation.

% \begin{acks}
% This work was supported in part by the National Natural Science Foundation of China (NSFC) under Grant No. U23B2031.The computations were performed on the High-Performance Computing (HPC) Platform of Hefei University of Technology.
% \end{acks}
\begin{acks}
This work was supported in part by the National Natural Science Foundation of China (NSFC) under Grant No. U23B2031. The authors gratefully acknowledge the High-Performance Computing (HPC) Platform of Hefei University of Technology for providing the computational resources that supported the experiments in this work.
\end{acks}

\clearpage
\balance
\bibliographystyle{ACM-Reference-Format}
\bibliography{sample-base}

\end{document}